\documentclass[prl,showpacs,amsmath,amssymb,twocolumn]{revtex4}
\usepackage{amsmath}
\usepackage{amsthm}
\usepackage{amssymb}
\usepackage{graphics}
\usepackage{feynmf}
\usepackage{xypic}
\input xy
\xyoption{all} \xyoption{arc}

\newcommand{\be}{\begin{equation}}
\newcommand{\ee}{\end{equation}}

\usepackage{bm}

\newcommand{\g}{{\mathfrak g}}
\newcommand{\legr}{{\mathfrak l}}
\newcommand{\rigr}{{\mathfrak r}}

\newcommand{\h}{{\mathcal{H}}}

\usepackage{bm}

\begin{document}

\title{On the Structure of QFT in the Particle Picture of the Path Integral Formulation}
\author{D.M. Jackson$^1$, A. Kempf$^2$, A. Morales$^3$}
\affiliation{$ $ \\ $^1$Department of Combinatorics and Optimization\\
$^2$Departments of Applied Mathematics and Physics\\
University of Waterloo, Ontario N2L 3G1, Canada\\
$^3$Department of Mathematics, MIT, Cambridge, MA, USA}

\begin{abstract}
In quantum field theory (QFT), the path integral is usually
formulated in the wave picture, \it i.e., \rm as a sum over field
evolutions. This path integral is difficult to define rigorously
because of analytic problems whose resolution may ultimately require
knowledge of non-perturbative or even Planck scale physics.
Alternatively, QFT can be formulated directly in the particle
picture, namely as a sum over all multi-particle paths, \it i.e.,
\rm over Feynman graphs. This path integral is well-defined, as a
map between rings of formal power series. This suggests a program
for determining which structures of QFT are provable for this path
integral and thus are combinatorial in nature, and which structures
are actually sensitive to analytic issues. For a start, we show that
the fact that the Legendre transform of the sum of connected graphs
yields the effective action is indeed combinatorial in nature and is
thus independent of analytic assumptions. Our proof also leads to
new methods for the efficient decomposition of Feynman graphs into
$n$-particle irreducible ($n$PI) subgraphs.
\end{abstract}

\maketitle

\noindent At the heart of the path integral formulation of quantum
field theory, \it e.g., \rm on flat space, is the integral over
fields,
 \begin{equation} Z[J] = \mu
\int e^{i S[\Phi] +i \int J\Phi~d^rx} D[\Phi] ~,\label{one}
\end{equation}
\it i.e., \rm the Fourier transform of $e^{iS}$. Here, $S$ is the
classical action, $\mu=e^{i\Lambda}$ corresponds to the cosmological
constant, $\Phi,J$ stand for (a collection of), \it e.g., \rm real
bosonic fields and their corresponding sources, and $c=\hbar=1$. We
assume suitable ultraviolet and infrared cutoffs so that the space
of fields, equipped with the inner product $\langle J,\Phi \rangle =
\int J(x) \Phi(x) d^rx$, is of finite dimension, say $N$. Choosing
an orthonormal basis, $\{b_a\}_{a=1}^N$, in the space of fields, we
have $\Phi = \Phi_a b_a,~J = J_a b_a,~ \langle J,\Phi\rangle
=J_a\Phi_a$, and:
\begin{equation} S[\Phi] = \sum_{n\ge 2} \frac{1}{n!}
S^{(n)}_{a_1,...,a_n}\Phi_{a_1}\cdots \Phi_{a_n}~.\label{actionexpansion}
\end{equation}
Twice occurring indices are to be contracted, {\it i.e.,} summed
over. $S^{(2)}$ is assumed to contain a Feynman $i\epsilon$ term,
and we assume $S$ does not lead to a non-zero vacuum field
expectation value. Then, $Z[J] = \mu \int_{\mathbb{R}^N} e^{iS[\Phi]
+iJ_a\Phi_a} \prod_j d\Phi_j$, after pulling the interaction terms
before the integral by using derivatives, completing the squares and
carrying out the integrations, reads:
\[ Z[J] = \mu'e^{\sum_{n>2}
\frac{i}{n!}S^{(n)}_{a_1,...,a_n}\partial_{(iJ_{a_1})}\cdots
\partial_{(iJ_{a_n})}} e^{(iJ_b)
\frac{i}{2}{S_{bc}^{(2)}}^{-1}(iJ_{c})}\] Thus, $Z[J]$ is the
generating functional of all Feynman graphs $\g$ built from the
Feynman rules $edge= i (S^{(2)})^{-1}$, and $n$-$vertex = iS^{(n)}$,
with at least one edge. We can view $Z[J]$ also as a sum of all
graphs with the additional Feynman rule $1$-$vertex= iJ_a$, where
each graph $\g$ has a symmetry factor $\omega(\g)=2^{-\ell}k^{-1}$.
Here, $\ell$ is the number of edges of $\g$ joining a vertex with
itself and $k$ is the number of automorphisms of $\g$. Note that if
$\g$ is a tree graph with labelled ends (\rm i.e., no $1$-vertices)
then $\omega(\g)=1$.


Correspondingly, let us denote the sum of only the connected graphs
by $iW[J]$. When exponentiated, it yields the sum of all graphs,
{\it i.e.,} $\exp(iW[J])=Z[J]$, as is easy to see combinatorially.
Further, as is well-known:

\bf Theorem 1: \rm \it For a given action, $S$, let $iW[J]$ denote
the sum of all connected graphs. Assume that the power series $W[J]$
converges to a function which is convex. The definition
$\varphi_a=\partial W[J]/\partial J_a$ can, therefore, be inverted
to obtain $J[\varphi]_a=(J[\varphi])_a$. Then, the Legendre
transform of $W[J]$, namely
$ 
\Gamma[\varphi] = -  J[\varphi]_a\varphi_a + W[J[\varphi]]
$, 
yields $i\Gamma[\varphi]$, which is the generating functional of the
sum of $n$-point $1$-particle irreducible (1PI) graphs for $n>2$,
and $i\Gamma^{(2)} = iS^{(2)} +$ $\sum$ ($2$-point 1PI graphs). \rm
Thus, overall:
$$
\xymatrixcolsep{3pc} \xymatrix{ e^{iS[\Phi]} \ar[r]^{Fourier} & Z[J]
\ar[l]\ar[r]^{log/exp} & iW[J] \ar[l]\ar[r]^{Legendre} & \ar[l]
i\Gamma[\varphi]. }
$$
We will here question the assumptions underlying Thm.1. First, let
us consider Theorem 1's broad significance:

\bf A) \rm \it The practical calculation of Feynman graphs. \rm Any
connected graph can be viewed as consisting of maximal 1PI subgraphs
that are connected by edges whose deletion would disconnect the
graph. For practical calculations of Feynman graphs, this
conveniently identifies the 1PI graphs as building blocks. After
renormalizing them, the 1PI graphs may be glued together to form
connected graphs with no further loop integrations needed. Later we
shall discuss a strategy for the further decomposition of 1PI graphs
by extending Thm.1 through higher order, \it i.e., \rm multi-field
Legendre transforms.

\bf B) \rm \it The action and the generating functional of tree
graphs are related by Legendre transform. \rm Continuing the
discussion of the structure of connected graphs in (\bf A\rm), we
notice that
any connected graph is a tree graph whose vertices are $1$PI graphs
connected by strings of edges and $2$-point $1$PI graphs. Thus, the
sum of connected graphs, $iW[J]$, is also the sum of all tree graphs
made from new Feynman rules. The new Feynman rules' $n$-vertex is
the sum of all $n$-point $1$PI graphs, while the new rules' edge is
given by $
-~+~-\!\!\bigcirc\!\!-~+~-\!\!\bigcirc\!\!-\!\!\bigcirc\!\!-
~+~-\!\!\bigcirc\!\!-\!\!\bigcirc\!\!-\!\!\bigcirc\!\!- ~+~ \dots$
$=\left((-)^{-1} + \bigcirc\right)^{-1}$, where $\bigcirc$ is the
sum of $2$-point $1$PI graphs and we summed a geometric series.
Clearly, these are the Feynman rules generated by $\Gamma[\varphi]$
if viewed as an action. Thus, the Legendre transform maps the sum of
trees, $W$, into the action $\Gamma$. Now every generating
functional of tree graphs, $iT[K]$, is the generating functional of
connected graphs for some action, $F[\Psi]$, since Fourier transform
and exponentiation are invertible. Thus:

\it Theorem 1 (2nd formulation):  Let $iT[K]$ denote a sum of all
tree graphs built from the Feynman rules of some action, $F[\Psi]$.
Assume that the power series $T[K]$ converges to a function which is
convex, so that the definition $\Psi_a=\partial T[K]/\partial K_a$
is invertible, to obtain $K[\Psi]_a$. Then, $F[\Psi]$ and $T[K]$ are
related by Legendre transform:
\begin{equation}
F[\Psi] = -K[\Psi]_a\Psi_a + T[K[\Psi]]. \label{2nd}
\end{equation}
\rm Below, we will prove Thm.1 in this formulation but with weaker
assumptions.

\bf C) \rm \it The perturbative solution to the classical equations
of motion can be obtained from the sum of the tree graphs. \rm To
see this, consider the action, $F[\Psi]+\int K\Psi d^rx$, of a
classical system coupled linearly to a source field, or driving
force, $K$. The equations of motion, $\delta F/\delta\Psi = -K$, are
to be solved for the field $\Psi[K]$ as a functional of the applied
source $K$. By Thm.1 (2nd formulation), the inverse Legendre
transform $T[K]=F[\Psi]+\int K\Psi d^rx$ of $F[\Psi]$ yields the
generating functional, $iT[K]$, of trees. {}From the properties of
Legendre transforms we have: $\Psi[K]= \delta T[K]/\delta K$. Thus,
$-i\times$ the sum of the tree graphs, $iT[K]$, once differentiated
by $K$, yields the perturbative solution to the classical equations
of motion in powers of the perturbing source field $K$.

\bf D) \rm \it Effective action. \rm $\Gamma[\varphi]$ plays the
r\^{o}le of a quantum effective action because it is that action
which when treated classically yields the correct quantum theoretic
answer: any $n$-point function can be calculated as a sum of all
connected graphs using the Feynman rules of the action $S$ or also,
as if classical, \it i.e., \rm as a sum of all tree graphs only,
when adopting the Feynman rules generated by the effective action
$\Gamma$.


\bf E) \rm \it Duality of problems and solutions. \rm One usually
defines a problem by specifying an action, $S$, and the classical
and quantum solutions are then obtained by calculating $T$ and $W$
respectively. $T$ and $e^{iW}$ are the Legendre and Fourier
transforms of $S$ and $e^{iS}$, respectively. Both transforms are
invertible. Thus, one may also define a problem by specifying, say,
$T$ (or $W$). The problem's solution is then the action, $S$. In
fact, since the Legendre and Fourier transforms are involutive (up
to a trivial sign), $S$ can be calculated in the same way by using
new, ``dual" Feynman rules: a given $W$ is viewed as an action, the
dual Feynman rules are read off, and $iS$ is obtained as the sum of
all connected graphs. This duality was first noticed in the context
of statistical physics, in \cite{englert}. Here we add that,
similarly, a given $T$ can be viewed as an action, dual Feynman
rules can be read off, and $S$ can then be calculated from the sum
of all tree graphs. For example, in cosmology, there are efforts to
reconstruct the potential in the inflaton action, $S$, from the
inflaton correlation functions in $W$ obtained \it via \rm
measurements of the cosmic microwave background \cite{kinney}. In
principle, it should be possible to view $W$ (as far as $W$ is
known) as an action, read off the dual Feynman rules and calculate
$iS$ as the sum of connected graphs. Similarly, by summing up only
the tree graphs, one should, in principle, obtain the inflaton's
quantum effective action.

It appears that every theory specified by an action possesses a
Fourier dual as well as a Legendre dual theory. Thus, \it e.g., \rm
in addition to the Dyson-Schwinger equation $(\delta
S/\delta\Phi[-i\delta/\delta J]+J)e^{iW[J]}=0$, the involutive
property of the Fourier transform implies a dual Dyson-Schwinger
equation: $(\delta W/\delta J[i\delta/\delta \Phi]
-\Phi)e^{iS[\Phi]}=0$. Similarly, there are, \it e.g., \rm dual
Slavnov-Taylor identities for gauge theories. We note that an
instance where the Legendre transform of an effective action is
itself the effective action of a known theory was found in
\cite{ford} in the context of S-duality and weak versus strong
coupling regimes. Finally, we notice that the involutive property of
the Legendre transform implies that the ``sum of the trees of trees"
must reduce to the original sum of the Feynman rules. The involutive
property of the Fourier transform implies a corresponding statement
for connected graphs.


\bf From the wave picture to the particle picture. \rm All of the
above considerations appear to hinge on analytic assumptions.
Namely, it appears that $S,Z, W$ and $\Gamma$ should be series that
converge to well-defined functions which possess Fourier and
Legendre transforms, respectively. For example, the power series
$W$ and $\Gamma$  would seem to have to converge to convex functions
in order to possess Legendre transforms. As is well known, however,
not even their convergence can be assumed in QFT, a problem whose
solution, it is thought, may require knowledge of non-perturbative
or even Planck scale physics.

The fact that perturbative QFT is nevertheless very successful in practice
suggests that it should be possible to make the formalism of QFT
mathematically well-defined without analytic assumptions such as convexity
or even convergence. Within such a framework, it should be possible to
prove key theorems combinatorially, such as Thm.1, the involution properties of
the Fourier and Legendre transforms, or the Dyson-Schwinger equations.

To this end, we define $S,Z,W,\Gamma, T$ and $F$ as elements in a
ring of formal power series, for bosons as for fermions. All
physically relevant information is encoded in the individual
coefficients. For rings of formal power series, see, \it e.g., \rm
\cite{jackson}. For any formal power series $F$, with $F^{(1)} =0$
and $F^{(2)}$ invertible, (which would require only local
convexity), we then define a ``combinatorial Legendre transform",
$T$, namely as the following map: view $F[\Psi]$ as an action, read
off the Feynman rules and then obtain $iT[K]$ as the power series
generating all tree graphs. We also define a ``combinatorial Fourier
transform", $e^{iW[J,z]}$ of $e^{iS[\Phi]}$: read off Feynman rules
from $z^{-1}S$, where $z$ is an indeterminate, and set $iW = z\sum
connected~graphs$ with the combinatorial factors $\omega(\g)$. While
the edge and vertices are proportional to $z$ and $z^{-1}$, no
negative powers of $z$ occur in $W$. This is because for any
connected graph, $\g$, the numbers of edges and vertices, $E(\g)$
and $V(\g)$, obey $E(\g)-V(\g)\ge -1$. Note that $z$ counts powers
in $\hbar$ (and thus loops). Indeed, the combinatorial Legendre
transform
is contained in $iW[J,z]$ as the term proportional to $z^0$. This is
because exactly for tree graphs, as is easy to verify:
\begin{equation}
1 = V(\g)-E(\g). ~\label{ec}
\end{equation}
Within this framework, $Z$ is defined not through Eq.\ref{one}, \it
i.e., \rm  as a sum over all field evolutions (the wave picture) but
instead through $Z=e^{iW}$ as a sum over all multi-particle paths
(the ``particle picture"), where the term ``path" means graph.
Notice that the principle that a particle's classical path is, in a
suitable measure, the shortest path, while quantum theory requires a
sum over all paths, persists in second quantization: while the
classical solutions are obtained from the tree graphs only, QFT
requires summing over all graphs. Indeed, tree graphs are the
shortest graphs in terms of the number of edges for any given number
of leaves of the graph, {\it i.e.,} for any given perturbation
order. Also, the free propagator, {\it i.e.,} the edge, can itself
be viewed as a sum over paths. A ``path" in QFT is, then, a graph of
paths.

The QFT path integral is mathematically well-defined through the
combinatorial Fourier transform because the calculation of each
coefficient involves only a finite number of terms. This suggests the
program of trying to prove key equations of QFT combinatorially, for
example the Dyson-Schwinger equations, or Eq.\ref{2nd}. This is
non-trivial because, where successful, it shows that the equation in
question is fundamentally combinatorial in nature and does not hinge upon
analytic assumptions - such as assumptions of convergence and convexity in
Eq.\ref{2nd} of Thm.1, or, in the case of the Dyson-Schwinger equations,
the assumption that boundary terms can be neglected when path integrating
a total derivative. While one aim is to reveal the robustness or fragility
of the key equations of QFT with respect to analytic assumptions, any
deeper understanding of the key equations in QFT has of course the
potential to reveal useful new structures.

Starting this program, we here give a transparent and bare-bones
combinatorial proof of Thm.1 which shows that the theorem is robust
against issues of analyticity. Our proof shows that the Legendre
transform in QFT can be understood, more deeply, as a simple statement
(namely Eq.\ref{ec}) about tree graphs. This insight then leads to useful
new results, namely about the decomposition of Feynman graphs into their
$n$PI components.

\bf Theorem 1 in the new framework. \rm In the second formulation of Thm.1
above, $F$ may or may not be an effective action. Our aim is to prove
Thm.1 in this general form, but for the combinatorial Legendre transform.

\it Theorem 1 (3rd, combinatorial formulation):  Let $F[\Psi] =
\sum_{n\ge 2} \frac{1}{n!} F^{(n)}_{a_1,...,a_n}\Psi_{a_1}\cdots
\Psi_{a_n}$ be an element of a ring of formal power series in
commutative indeterminates $\Psi_a$. Assuming that the coefficient
matrix $F^{(2)}$ is invertible, $F[\Psi]$ can be viewed as an action
that defines Feynman rules. The sum of their tree graphs yields a
formal power series, say $iT[K]$, in variables $K_a$.
By definition, we relate the variables $\Psi$ and $K$ through the
algebraic derivative $ \Psi[K]_a = \partial T[K]/\partial K_a,
\label{psi}$ which is a well-defined operation in the ring, so that
$\Psi[K]_a$ is a formal power series in the $K_b$. Then, the formal
power series $F[\Psi]$ and $T[K]$ obey the Legendre transform
equation:
\begin{equation}
T[K] = K_a\Psi[K]_a  + F[\Psi[K]]. \label{toshow}
\end{equation} \rm
We remark  that $F[\Psi[K]]$ is well-defined as a formal power series
since $\Psi[K]$ has no constant term. The theorem covers the special case
when the usual analytic Legendre transform of $T[K]$ is well-defined, \it
i.e., \rm the case in which $T[K]$ obeys the analytic conditions of Thm.1
in its second formulation. This is because in this case the transformed
variable obeys $\Psi^{analytic}_a =
\partial T[K]/\partial K_a = \Psi_a$ and therefore with $F^{analytic} := -
\Psi_a K_a +T$ and Eq.\ref{toshow} we have $F=F^{analytic}$. We can then
conclude that the Feynman rules underlying $T$ are generated by the
Legendre transform $F^{analytic}$. Notice that Thm.1 in its 1st and 2nd
formulations makes no claim when the sum of the tree graphs does not
converge to a function, or does converge but the function is not convex.
Our generalized Thm.1 (3rd formulation) shows that Eq.\ref{toshow} holds
even then.

In the literature, Thm.1 is proven in the first formulation above, see
\cite{dominicis,jonadewitt,vasiletal}. The proof by Weinberg,
\cite{WeinVol2}, essentially addresses Thm.1 in its second formulation,
{\it i.e.,} directly as a map between any action and its sum of tree
graphs. However, that proof relies on analytic assumptions and requires
the taking of a subtle limit.

\bf Combinatorial proof of Thm.1. \rm Our proof strategy is to show that
$i\times$Eq.\ref{toshow} is term by term equivalent to a much simpler
statement, namely Eq.\ref{ec}. To this end, let us prove the power series
equation $i\times$Eq.\ref{toshow} for the coefficients of each $m$-power of $K$ for $m\geq 2$.
That is,
\begin{eqnarray} \label{eqfivep}
\lefteqn{\partial/\partial_{iK_{a_1}}\dots\partial/\partial_{iK_{a_m}}
iT\vert_{K=0} = }\\
&&
\partial/\partial_{iK_{a_1}}\dots\partial/\partial_{iK_{a_m}}
\left(iK_a\Psi[K]_a  + iF[\Psi[K]]\right) \vert_{K=0}. \nonumber
\end{eqnarray}
By definition of $T[K]$, on the LHS of Eq.\ref{eqfivep} we obtain
the sum of all tree graphs $\g$  with $m$ ends, labelled by
$a_1,\dots,a_m$, with each such $\g$ occurring exactly once and with
$\omega(\g)=1$. We will complete the proof by showing that the RHS
consists of all such graphs with multiplicity $V(\g)-E(\g)$. We
begin by writing the RHS of $i\times$Eq.\ref{toshow} in terms of
tree graphs. We have  $iK_a=$~$1$-$vertex$,
 and
\begin{equation}
iF[\Psi[K]] = \sum_{n\ge 2} \frac{i}{n!}
F^{(n)}_{a_1,...,a_n}\Psi[K]_{a_1}\cdots \Psi[K]_{a_n} \label{F},
\end{equation}
contains $iF^{(2)}_{a_1,a_2}= -(edge)^{-1}$ and
$iF^{(n)}_{a_1,...,a_n}=n$-$vertex$ for $n>2$. Thus, the RHS of
$i\times$Eq.\ref{toshow} takes the form:
\begin{eqnarray*}
\lefteqn{(1\mbox{-}vertex)_a\Psi[K]_{a}-\frac{1}{2}\Psi[K]_{b_1}(edge)_{b_1,b_2}^{-1}\Psi[K]_{b_2} +}\\
&& \sum_{n>2} \frac{1}{n!}
(n\mbox{-}vertex)_{a_1\cdots a_n}\Psi[K]_{a_1}\cdots\Psi[K]_{a_n}.
\end{eqnarray*}
After differentiating $m$ times and setting $K=0$,
we obtain Eq.6, which contains only graphs with $m$ labelled ends.
  Since, by definition, $\Psi[K]_b=\partial
iT[K]/\partial(i K_b)$ is the sum of trees with one end vertex
removed, the RHS of Eq.\ref{eqfivep} is the sum of all tree graphs
with $m$ labelled ends obtained by taking a term of the action, \it
i.e., \rm $-(\mbox{edge})^{-1}$ or any $n$-vertex, and attaching the
sum of all tree graphs at each of its free indices. After
simplification, this means that Eq.5 reads, schematically:
\[
\xy 0;/r.20pc/: (-11.25,0)*\ellipse(7,3.5)__,=:a(360){.};
(-19.5,0)*{{\scriptstyle  \sum \text{trees}}~ = }; (-2,0)*{ \xy
(0,-3.5)*{}="A"; (0,-6.5)*{}="B"; "B"*{\scriptstyle \bullet};
"A";"B"**\dir{-}; (0,0)*{\scriptstyle \sum \text{trees}};
(0,0)*\ellipse(7,3.5)__,=:a(360){.};
\endxy
}; (9,0)*{-}; (20,0)*{ \xy (-4,0)*{ \frac{1}{2!} };
(3,7)*{\scriptstyle \sum \text{trees}};
(1.5,3.5)*\ellipse(7,3.5)__,=:a(360){.}; (4,0)*{ \xy 0;/r.18pc/:
(-2,0)*{}="AA"; (0.5,-0.5)*{}="A"; (0.5,5.5)*{}="B"; (5,3.5)*{}="C";
"A"; "B" **\dir{-};
\endxy
 };
(1.5,-3)*\ellipse(7,3.5)__,=:a(360){.}; (3,-6)*{\scriptstyle \sum
\text{trees}};
\endxy};
(32,0)*{+}; (40,0)*{ \frac{1}{3!} }; (49,1)*{ \xy
 (0,0)*{\xy (2,-1)*{}="A"; (6,-1)*{}="B";
(4,-4)*{}="C"; (4,-7)*{}="D"; "A"; "C"
 **\dir{-}; "B"; "C" **\dir{-}; "D"; "C" **\dir{-};
"C"*{\scriptstyle \bullet};
\endxy
}; (0.2,-3.2)*\ellipse(7,3.5)__,=:a(360){.};
(-3.8,2.8)*\ellipse(7,3.5)__,=:a(360){.};
(3.8,2.8)*\ellipse(7,3.5)__,=:a(360){.}; (0.2,-6.5)*{\scriptstyle
\sum \text{trees} }; (-7.8,5.8)*{\scriptstyle \sum \text{trees} };
(7.8,5.8)*{\scriptstyle \sum \text{trees} };
\endxy};
(65,0)*{+ \cdots};
\endxy
\]
Let us consider an arbitrary tree graph, $\g$, with $m$ labelled
ends. On the LHS, it occurs exactly once. To count its occurrences
on the RHS, we choose an arbitrary edge $e$ of $\g$, and let $\legr$
and $\rigr$ arbitrarily denote the two subtrees to either side of
the edge. In the second term on the RHS, the edge $e$ occurs twice,
and because of the $1/2$ in the action, $g$ occurs with weight
$-E(\g)$.

We now choose an arbitrary $n$-vertex $v$ of $\g$, where
$n\in\{1,3,4,...\}$. Let $\{t_j\}_{j=1}^n$ arbitrarily denote the
sub-trees emanating from its legs. From the remaining terms on the
RHS of Eq.\ref{toshow}, our vertex $v$ with the attached subtrees
$\{t_j\}_{j=1}^n$ arises $n!$ times, which is cancelled by the
$1/n!$ in the action. Thus, $\g$ occurs $V(\g)$ times in the
remaining terms and therefore indeed with overall weight
$V(\g)-E(\g)$ on the RHS.

\bf Outlook. \rm In a follow-up paper, we will show how insight from our
combinatorial proof of Thm.1 yields a strategy for a more efficient
decomposition of Feynman graphs into $n$PI graphs for practical
calculations. Namely, recall that we reduced the Legendre transform,
Eq.\ref{toshow}, to the simple equation Eq.\ref{ec}, which is Euler's
formula, a special case of the general Euler-Poincar\'e formula for the
homology of graphs. As we will show in the follow-up paper, the so-called
cactus representation $\h(\g)$ of $n$PI graphs $\g$, see \cite{cactus},
can be used to obtain a generalization of Eq.\ref{ec} to $1=   V(\h(\g))-
E(\h(\g)) +C(\h(\g))$ (where $C$ is the number of cycles) which allows us
to generalize the diagrammatic analysis of \cite{dominicis} and improve on
the results on higher order Legendre transforms and the analysis of
Dyson-Schwinger equations of \cite{vasiletal}. We will also show that the
involutive property of the Legendre transform can be proven purely
combinatorially. This means that a) the sum of trees of trees indeed
always reduces to the original sum of Feynman rules and b) that every
theory has a Legendre dual whose dual is the original theory. We will also
study the involutive property of the Fourier transform.



There are a number of further key equations of QFT which are usually
derived using analytic arguments and assumptions, and it should be
very interesting to use the combinatorial Fourier transform to
investigate which of these equations are actually of a purely
combinatorial nature and therefore robust against analytic
difficulties. For example, the origins of anomalies and of ghost
fields are usually traced, analytically, to the measure in the wave
picture path integral, Eq.\ref{one}. Is there a combinatorial
derivation of these, perhaps involving what could be viewed as a
``measure" in the particle picture path integral? The usual
derivation of the Dyson Schwinger equations assumes that boundary
terms in Eq.\ref{one} can be neglected. What is the combinatorial
analog of derivatives and boundary terms for the combinatorial
Fourier transform of the particle picture path integral? Also,
certain discontinuities in QFT renormalization can be traced to
functional analytic restrictions of the domain of the wave picture
path integration, see \it e.g., \rm \cite{klauder}. What is the
analog in the particle picture path integral?
\smallskip\newline\noindent \bf Acknowledgement. \rm We acknowledge
support by CFI, OIT, and the Discovery and CRC programs of NSERC.



\begin{thebibliography}{11}

\bibitem{englert} C. deDominicis, F. Englert, J.Math.Phys., {\bf 8},
2143 (1967)

\bibitem{kinney} W.H. Kinney, E.W. Kolb, A. Melchiorri, A. Riotto,
Phys. Rev. {\bf D74}, 023502 (2006)

\bibitem{ford} C. Ford, I. Sachs, Phys. Lett. {\bf B362}, 88 (1995)


\bibitem{jackson} I.P. Goulden, D.M. Jackson, \it Combinatorial Enumeration,
\rm Dover, Mineola N.Y. (1983)


\bibitem{jonadewitt}
G. Jona-Lasinio, Nuovo Cim., {\bf 34}, 1790 (1964)

\bibitem{dominicis}
C. deDominicis, P.C. Martin, J. Math. Phys., {\bf 5}, 14, (1964), C.
deDominicis, P.C. Martin, J. Math. Phys., {\bf 5}, 31, (1964)

\bibitem{vasiletal}
A.N. Vasil'ev, \it Functional Methods in Quantum Field Theory and
Statistical Physics, \rm Gordon \& Breach, N.Y. (1998), A.N.
Vasil'ev, A.K. Kazanskii, Theor. Math. Phys., {\bf 12}, 875 (1972),
Yu.M. Pis'mak, Theor. Math. Phys. {\bf 18}, 211 (1974)

\bibitem{WeinVol2}
S. Weinberg, \it The Quantum Theory of Fields II, \rm CUP,
Cambridge, U.K., (1996)

\bibitem{cactus}
E.A. Dinits, A.V. Karzanov, M.V. Lomonosov, Stud. Disc. Opt., 290 (1976),
D. Naor, V.V. Vazirani, Proc. WADS, 273 (1991), Y. Dinitz, Z. Nutov, Proc.
STOC, 509 (1995)

\bibitem{klauder} J.R. Klauder, \it Beyond Conventional Quantization, \rm
CUP, Cambridge, U.K. (2000)
\end{thebibliography}
\end{document}